\documentstyle[12pt]{article}
\def\ds{\displaystyle}
\newcommand \dsty \displaystyle
\newcommand{\beq}{\begin{equation}}
\newcommand{\eeq}{\end{equation}}
\newcommand{\bea}{\begin{eqnarray}}
\newcommand{\eea}{\end{eqnarray}}
\newcommand{\Lam}{\Lambda_{\overline{MS}}}

\newcommand{\MS}{\overline{MS}}
\newcommand{\nn}{\nonumber}
\begin{document}
\setlength{\baselineskip}{16pt}
\setcounter{page}{0}
\begin{titlepage}
\begin{center}
                      \hfill CERN--TH/96-193 \\
                     \hfill PM/95-27  \\       
                     \hfill IMPERIAL TP 95-96/08 \\
                      \hfill   hep-ph/yymmxxx \\
 
\vskip .25in
{\Large \bf
               Chiral Symmetry Breaking in QCD: \\
A Variational Approach\\
}
C. Arvanitis$^\dagger$ \footnote{supported by the EC
under H.C.M Grant No ERBCHBICT941235},
F. Geniet$^\ddagger$, 
J.-L. Kneur$^{\star}$\footnote{On leave
from Laboratoire de Physique
Math\'ematique, Universit\'e  
Montpellier II-CNRS,
34095 Montpellier c\'edex 05.}
and  
A. Neveu$^\ddagger$
\\
{$^\dagger$ \em Physics Department,  
Imperial College,  \\ 
Theoretical Physics Group, \\ 
Prince Consort Rd., 
 London SW7 2BZ, UK} \\
\vskip .3 cm
{$^\ddagger$ \em Laboratoire de Physique Math\'ematique\footnote{Laboratoire
associ\'e au Centre National de la Recherche Scientifique.}
\\
Universit\'e Montpellier II-CNRS, 
F-34095 Montpellier cedex 05.}\\
\vskip .3 cm
{$^{\star}$ CERN,
\em Theoretical Physics Division \\
CH-1211 Geneva 23 Switzerland} \\
\end{center}

\begin{abstract}
We develop a ``variational mass" expansion approach, 
recently introduced in the 
Gross--Neveu model, to evaluate 
some of the order parameters 
of 
chiral
symmetry breakdown in QCD. The method relies on a reorganization of
the usual perturbation theory with the addition of 
an {\em arbitrary} quark mass $m$,
whose 
non-perturbative behaviour is inferred partly  
from renormalization group properties, 
and from analytic continuation in $m$ properties.
The resulting ansatz 
can be optimized, 
and in the chiral limit $m \to 0$ 
we 
estimate
the  
dynamical contribution to the 
``constituent" masses of the light quarks $M_{u,d,s}$; 
the pion decay
constant $F_\pi$ and the quark condensate
$\langle \bar q q \rangle$. 
\end{abstract}
\vskip 3 cm
CERN--TH/96-193 \\
August 1996
\end{titlepage}
\setlength{\baselineskip}{15pt}
\section{Introduction}
\setcounter{footnote}{0}
A still challenging question in strong interaction physics is 
the derivation of the 
low-energy properties of the spectrum from ``QCD first principle",
due to our limited present skill with non-perturbative physics.
At very low energy, where the ordinary 
perturbation theory cannot be applied, 
Chiral Perturbation Theory~\cite{GaLeut} and Extended Nambu--Jona-Lasinio
models~\cite{NJL,ENJL}~\footnote{For a recent complete review see 
\cite{Miransky}.} give a consistent framework
in terms of a set of parameters that have to be fixed from the data;
yet the bridge between those effective parameters and the basic
QCD degrees of freedom remains largely unsolved. Although lattice QCD 
simulations
recently made definite progress~\cite{lattcsb} in that direction, 
the consistent treatment of dynamical
unquenched quarks and the chiral symmetry remains a serious problem. \\

In this paper, 
we investigate a new, semi-analytical method, to 
explore {\it how far}
the basic QCD Lagrangian can provide, in a self-consistent 
way, 
non-zero 
dynamical quark
masses, quark condensates, and pion decay constant, 
in the limit of vanishing Lagrangian (current) quark masses. Such a 
qualitative picture
of chiral symmetry breakdown (CSB) can be made more quantitative
by applying
a new ``variational mass" approach, 
recently developed within the framework of the
anharmonic oscillator \cite{bgn}, and in  
the Gross--Neveu (GN) model \cite{gn1,gn2}.
The starting point is very similar to
the ideas developed a long time ago and implemented 
in various different forms in refs.\cite{pms}, \cite{delta}. 
There, it was advocated 
that the convergence of conventional perturbation
theory may be improved by a variational procedure in which
the separation of the
action into ``free" and ``interaction" parts
is made to depend on some set of auxiliary parameters.
The results obtained  by expanding to finite order
 in this redefined perturbation
series are optimal in regions  of the  space of auxiliary
parameters where they are least sensitive to these parameters.
Recently there appeared
strong evidence that this optimized perturbation
theory may indeed lead to a rigorously convergent
series of approximations even in strong coupling cases
\cite{JONE}. \\
An essential novelty, 
however, in \cite{bgn}--\cite{gn2} and the present paper,  
is that our construction combines in a specific manner 
the renormalization group (RG) invariance  
with the properties of an analytically continued, arbitrary mass parameter
$m$.  
This, at least in a certain approximation to be motivated, 
allows us to reach 
{\it infinite} order of 
the variational-perturbative expansion, and therefore presumably optimal,
provided it converges. Our main results are a set of
non-perturbative
ansatzs for the relevant CSB quantities, 
as functions of the variational mass $m$, 
which can be 
studied for extrema and optimized. 
Quite essentially, our construction also provides a simple and
consistent treatment of the renormalization, reconciling 
the variational approach with the inherent infinities
of quantum field theory and the RG properties.

Before proceeding, let us note that there exists
a quite radically different attitude towards CSB in QCD, advocating 
that the responsible mechanism is most probably the non-perturbative
effects due to the {\em instanton} vacuum~\cite{Callan},
or even more directly related to confinement~\cite{Cornwall}.
However, 
even if the instanton picture of CSB is on general grounds well motivated,
and many fruitful ideas have been developed in that 
context\footnote{See e.g 
ref.~\cite{Shuryak} for a review and original references.}, 
as far as we are aware there is at present no 
sufficiently rigorous or compelling evidence for it, derived from ``first
principle".  
In any event, it is certainly of interest to 
investigate quantitatively 
the ``non-instantonic" contribution to CSB, and we hope  
that our method is a more consistent step in that direction.
\section{Dynamical quark masses}
In what follows we only consider the $SU(n_f)_L \times SU(n_f)_R$
part of the chiral symmetry, realized by the QCD Lagrangian 
in the absence of quark mass terms, 
and for
$n_f =2$ or $n_f = 3$ as physically relevant applications.

Following the treatment of the anharmonic oscillator~\cite{bgn} and its
generalization to the GN model~\cite{gn1,gn2}, 
let us consider the following modification of the usual QCD Lagrangian, 
\beq
L_{QCD} \to L^{free}_{QCD}(g_0 =0, m_0=0)
-m_0 \sum^{n_f}_{i=1} \overline{q}_i q_i
+ L^{int}_{QCD}(g^2_0 \to x g^2_0)
+x \; m_0 \sum^{n_f}_{i=1} \overline{q}_i q_i\;,
\label{xdef}
\eeq
where $L^{int}_{QCD}$ designates the ordinary QCD interaction terms, and
$x$ is a convenient ``interpolating" expansion parameter.
This formally is equivalent 
to substituting everywhere in the bare Lagrangian,  
\beq m_0 \to m_0\; (1-x); ~~~~g^2_0 \to g^2_0\; x, 
\label{substitution}
\eeq
and therefore 
in any perturbative (bare) quantity as well, 
calculated in terms of $m_0$ and $g^2_0$.
Since the original massless QCD
Lagrangian is recovered for $x \to 1$, $m_0$ is to be considered as 
an {\it arbitrary } mass parameter 
after substitution (\ref{substitution}). One expects
to optimize physical quantities with
respect to $m_0$ at different, possibly arbitrary 
orders of the expansion parameter $x$,
eventually approaching a stable limit, i.e {\it flattest}
 with respect to $m_0$,
 at sufficiently high order in $x$. \\
However,
before accessing any physical quantity
of interest for such an optimization, the theory should be
renormalized, and there is an unavoidable mismatch
between the expansion in $x$, as
introduced above, and the ordinary perturbative expansion as dictated by the
mass and coupling counterterms.
Moreover, 
it is easy to see that 
at any finite order in the $x$ expansion,
one always 
recovers a trivial result in the limit $ m \to 0$ (equivalently 
$x\to 1$), which
is the limit 
in which to identify non-zero 
order parameters of CSB. 

These 
problems can be circumvented by 
advocating a specific ansatz, which resums 
the (reorganized) perturbation series in $x$ and is such that
the limit $x \to 1$ 
no longer gives
a trivial zero mass gap. 
As was shown in detail in ref.~\cite{gn2}, the ansatz for the dynamical
mass is
most easily derived by following the steps\footnote{See 
also ref.~\cite{jlk} for
a detailed derivation in the QCD context.}:\\

{\it i}) Consider first the general
solution for the running mass,
given as
\beq
m(\mu^{'}) = m(\mu )\;\; {\exp\left\{ -\int^{g(\mu^{'})}_{g(\mu )}
dg {\gamma_m(g) \over {\beta(g)}} \right\} }
\label{runmass}
\eeq
in terms of the effective coupling $g(\mu)$, whose RG evolution is given
as $\mu dg(\mu)/d\mu \equiv \beta(g)$, and $\gamma_m(g) \equiv 
-(\mu/m)d(m(\mu))/d\mu$.
Solving (\ref{runmass}) imposing the 
``fixed point" boundary condition:
\beq
M \equiv m(M),
\label{RGBC}
\eeq
at two-loop RG order we obtain, after some algebra 
(we use the normalization 
$\beta(g) = -b_0 g^3 -b_1 g^5 -\cdots$,
$\gamma_m(g) = \gamma_0 g^2 +\gamma_1 g^4
+\cdots$):
\beq
M_2 =  \bar m \;\;
\ds{f^{-\frac{ \gamma_0}{2b_0}}\;\; \Bigl[\frac{ 1 +\frac{b_1}{b_0}
\bar g^2 f^{-1}}{ 1+\frac{b_1}{b_0}\bar g^2} \Bigr]^{ -\frac{\gamma_1}{
2 b_1}
+\frac{\gamma_0}{2 b_0}   } }\;,
\label{MRG2}
\eeq
where
$\bar m \equiv m(\bar\mu)$, $\bar g \equiv g(\bar\mu)$ 
($\bar \mu \equiv \mu \sqrt{4 \pi} e^{-\gamma_E/2}$), and
$f \equiv \bar g^2/g^2(M_2)$ satisfies
\beq
f = \ds{ 1 +2b_0 \bar g^2 \ln \frac{M_2}{\bar \mu }
+\frac{b_1}{b_0} \bar g^2
\ln \Bigl[\frac{ 1 +\frac{b_1}{b_0} \bar g^2 f^{-1}}{
 1 +\frac{b_1}{b_0} \bar g^2 }\;f\;\Bigr] }\; ;
\label{f2def}
\eeq
(note in (\ref{MRG2}) and (\ref{f2def}) 
the recursivity in both $f$ and $M_2$).
The necessary 
non-logarithmic perturbative corrections to those pure RG results
are then consistently  
included as
\beq
M^P_2 \equiv M_2 \;\Bigl(1 +{2\over 3}\gamma_0 {\bar g^2\over f}
+{K \over{(4 \pi^2)^2}}{\bar g^4\over f^2}+{\cal O}(g^6)\;\Bigr)\;,
\label{Mpole}
\eeq
where
the complicated
two-loop coefficient $K$ was calculated
exactly in ref.~\cite{Gray}. 
Equation.~(\ref{Mpole}) defines the (infrared-convergent, gauge-invariant) 
pole mass~\cite{Tarrach} 
$M^P_2$, in terms of the 
$\MS$ mass at two-loop order,  
and can be 
shown~\cite{jlk} to resum the leading (LL)
{\it and} next-to-leading logarithmic (NLL) 
dependence in $\bar m$ to all
orders. \\
{\it ii})
Perform in expressions (\ref{MRG2}), (\ref{f2def}), (\ref{Mpole})
the substitution
$
\bar m \to \bar m v $, 
and integrate the resulting expression, denoted by $M^P_2(v)$,
according to  
\beq
\frac{1}{2i\pi} \;\oint \frac{dv}{v}\;e^v M^P_2(v)\;, 
\label{contgen}
\eeq
where the contour is around the negative real $v$ axis. \\

In \cite{gn2} it was shown that the previous steps correspond
(up to a specific renormalization scheme (RS) change, 
allowed on general grounds from RG properties) to a resummation
of the $x$ series as generated from the substitution
(\ref{substitution})\footnote{$v$ is related to the original
expansion parameter $x$ as $x = 1-v/q$, $q$ being the order of the expansion.}.
Moreover this is in fact the only way of rendering compatible the
above $x$ expansion and the ordinary perturbative one, thus obtaining
finite results.
Actually the resummation coincides with the exact result in the
large-$N$ limit of the GN model. Now, since the
summation can be formally
extended to arbitrary RG orders~\cite{gn2}, including consistently as many
arbitrary perturbative correction terms as known in a given theory, 
in the QCD case we make the assumption that it  
gives an adequate ``trial ansatz",
to be subsequently optimized in a way
to be specified next. 
After appropriate rescaling of the basic parameters, $\bar g$
and $\bar m$, by introducing the RG-invariant basic scale $\Lam$~
\cite{Lambda}
(at two-loop order), 
and the convenient scale-invariant dimensionless ``mass" parameter  
\beq
m''\equiv  \ds{(\frac{\bar m}{ \Lam}) \;
2^{C}\;[2b_0 \bar g^2]^{-\gamma_0/(2b_0)}
\;\left[1+\frac{b_1}{b_0}\bar g^2\right]^{
\gamma_0/(2 b_0)-\gamma_1/(2 b_1)}}
\; ,
\label{msec2def}
\eeq
we end up with the following dynamical mass ansatz:
\beq
{ M^P_2 (m^{''})\over \Lam}
 = {2^{-C} \over{2 i \pi}} \oint dy {e^{\;y/m^{''}}
\over{F^A(y) [C + F(y)]^B}} {\left(1 +{{\cal M}_{1}\over{F(y)}}
+{{\cal M}_{2}\over{F^2(y)}} \right)},
\label{contour7}
\eeq
where $y \equiv m'' v$, and 
$F$ is defined as  
\beq
F(y) \equiv \ln [y] -A \; \ln [F(y)] -(B-C)\; \ln [C +F(y)],
\label{Fdef}
\eeq
with $A =\gamma_1/(2 b_1)$, $B =\gamma_0/(2 b_0)-\gamma_1/(2 b_1)$, 
$C = b_1/(2b^2_0)$, in terms of the RG coefficients
\cite{betagamma}. 
Finally the perturbative corrections in (\ref{contour7}) 
are simply given as 
${\cal M}_{1} =(2/3)(\gamma_0/2b_0)$ and ${\cal M}_{2} = K/(2b_0)^2$. \\ 

Observe in 
fact that, were we in a simplified QCD world, where there would be
{\em no} non-logarithmic perturbative contributions (i.e. such that  
${\cal M}_{1} = {\cal M}_{2} = \cdots = 0$ in (\ref{contour7})),
the latter ansatz would 
then resums exactly the $x$ variational expansion. 
In that case, (\ref{contour7}) would have a very simple
behaviour near the origin $m'' \to 0$. Indeed, it is easy to see
that (\ref{Fdef}) admits an expansion 
$
F(y) \simeq C^{(B-C)/A}\;y^{1/A}$ for $y \to 0$, 
which immediately implies that (\ref{contour7}) would
give a simple pole at $y \to 0$, 
with a residue giving $M_2 = (2C)^{-C}\;\Lam $.
Moreover one can always 
choose
an appropriate renormalization scheme in which $
b_2$ and $\gamma_2$ are set to zero, as well as all
higher order coefficients, so that there are no other corrections
to the simple above relation. 

Now, in the realistic world, ${\cal M}_1$, ${\cal M}_2$, etc can
presumably not be neglected.
We can nevertheless expand 
(\ref{contour7}) near $m'' \to 0$ for any known 
non-zero ${\cal M}_{i}$,
using
\beq
\label{hankel}
\frac{1}{2i \pi} \oint dy e^{y/m^{''}}  y^\alpha  =
\frac{(m^{''})^{1+\alpha}}{\Gamma[-\alpha]}\; ,
\eeq
and the resulting Laurent expansion in $(m'')^{1/A}$ may be
analysed for extrema and optimized at different, in principle
arbitrary $(m'')^{1/A}$ orders. An important point, however, is
that the 
perturbative corrections do depend on the RS choice, as is well known. 
Since the pure RG behaviour in (\ref{contour7})  
already gives the order of magnitude, $M \simeq {\rm const} \times \Lam$, 
we can hope that 
a perturbative but optimized treatment of the remaining corrections
is justifed. 
In other words we shall perform an 
``optimized perturbation" with respect to $m''$
around the non-trivial fixed point
of the RG solution. \\ 
 
To take into account this RS freedom, we first introduce 
in (\ref{contour7}) an arbitrary scale parameter $a$, 
from $\bar \mu \to a\; \bar \mu$. 
Accordingly the 
perturbative coefficients ${\cal M}_{i}$ in (\ref{contour7}) take a 
logarithmic dependence in $a$, simply  
fixed order by order from (\ref{MRG2})--(\ref{Mpole}) 
and the requirement that (\ref{contour7})
differs from the original $\MS$ expression only by higher order terms.
The $a$-dependence will eventually exhibit a non-trivial extrema 
structure and we shall also  
optimize the result with respect to 
$a$\footnote{This procedure indeed gave very good results~\cite{gn2}
in the GN model, where in particular for low values of $N$ the
optimal values found, $a_{opt}$, are quite different from 1.}.
Actually 
there are other possible changes of renormalization prescriptions 
affecting expression (\ref{contour7}) in addition to the $a$ dependence, 
which may be 
taken into account as well. More precisely, the second coefficient
of $\gamma_m(g)$, $\gamma_1$, do depend on the RS choice,
even in MS schemes~\cite{Collins}. 
As it turns out, this additional RS freedom is  
very welcome in our case: in fact,  
the previous picture is  
invalidated, due to the occurence of extra branch cuts 
in the $y$ plane at $Re[y_{cut}] > 0$, 
as given by the zeros of $dy/dF$ from
(\ref{Fdef}) (in addition 
to the original cut on the negative real $y$ axis).
This prevents using 
the expansion near the origin, eq.~(\ref{hankel}), 
since it would lead to 
ambiguities of ${\cal O}(\exp(Re[y]/m''))$
for $m'' \to 0$\footnote{
The origin of those singularities is rather  
similar to the ambiguities related
to
renormalons~\cite{renormalons}.
An essential difference, however, is that the present
singularities occur in the analytic continuation of a mass parameter
rather than a coupling constant, and
it is possible to move those singularities away by an 
appropriate RS change, 
as we discuss next. See ref.~\cite{jlk} for an extended discussion.}.
The specific contour around the negative real axis
was suggested by the known properties of the large
$N$ limit of the GN model, and it is not surprising
if the analytic structure 
is more complicated in QCD.
However, the nice
point is that the actual position of those cuts
do depend on the RS, via
$A(\gamma_1)$ in (\ref{Fdef}).
Defining 
$
\gamma^{'}_1 \equiv \gamma_1 +\Delta \gamma_1$, 
we can choose 
$Re[y_{cut}] \simeq 0$ for $\Delta\gamma_1 \simeq$ 0.00437 (0.00267) for 
$n_f =$ 2 ($n_f =$ 3), respectively.
We therefore consider~\cite{jlk} the general RS change 
\beq
m' = \bar m\;(1+B_1 \bar g^2 +B_2 \bar g^4)\;;\;\;\;g^{'2} = \bar g^2\;
(1 +A_1  \bar g^2 +A_2 \bar g^4)\;
\label{RSchange}
\eeq
(implying $\Delta\gamma_1 = 2b_0 B_1 -\gamma_0 A_1$),
and optimize with respect
to this new arbitrariness\footnote{ 
We also impose a further RS choice,
$
b^{'}_2 = 0$, 
$\gamma^{'}_2 = 0$,  
which fixes $A_2$, $B_2$ in (\ref{RSchange}) and 
guarantees that our two-loop convention for $\Lam$
remains unaffected. Note, however, that  
(\ref{RSchange}) implies $\Lam \to \Lam \; \exp\{\frac{A_1}{2b_0}\}\equiv 
\Lambda'$. In what
follows we express the results in terms of the original $\Lam$.}. 

However one soon realizes that 
our extension of the ``principle of minimal sensitivity" (PMS)~\cite{pms}
defines a rather complicated optimization  
problem. 
Fortunately, we can study this problem within some approximations, 
which we believe are legitimate. 
Since 
the ansatz 
(\ref{contour7}) (with the above RS change understood, to make
it consistent) would indeed 
be optimal with respect
to $m^{''}$ for {\em vanishing} perturbative non-logarithmic 
corrections, ${\cal M}_{i} =0$,  
we shall assume that the 
expansion for small $m''$ is as close as possible to an optimum, 
and 
define  the $m^{''} \to 0$ limit by some relatively crude but standard 
approximation, 
avoiding numerical optimization
with respect to $m^{''}$. 
The approximation we are looking for is 
not unique: given (\ref{contour7}), one could construct  
different 
approximations leading to a finite limit
for $m'' \to 0$~\cite{gn2}. Here 
we shall only demonstrate
the feasibility of our program in the simplest possible
realization.  
In fact, since we shall anyhow optimize with respect to the RS dependence
we assume that it largely 
takes into
account this non-uniqueness due to higher order uncertainties.

Pad\'e approximants are known to greatly improve perturbative
results~\cite{pade}
and often have the effect of smoothing the RS dependence.  
We thus take a simple Pad\'e approximant
which by construction restitutes a simple pole for $F \to 0$ 
(i.e $m'' \to 0$) in 
(\ref{contour7}), and gives 
\beq
{M^{Pad\acute{e}}(a,\Delta\gamma_1,B_1,m''\to 0) = \Lam\;
(2C)^{-C} \;a\; \exp\{\frac{A_1}{2b_0}\}\;\left[
1 -{{\cal M}^2_{1}(a, \Delta\gamma_1, B_1)\over{{\cal M}_{2}(a, 
\Delta\gamma_1, B_1)}}\right] }
\label{Mpade}
\eeq
We have performed a rather systematic study of the possible extrema
of (\ref{Mpade}) for arbitrary 
$a$, $B_1$ (with $\Delta\gamma_1$ 
fixed so that the extra cuts start at $ Re[y] \simeq 0 $).
We obtain the flattest such extrema for $a \simeq 2.1$, $B_1 \simeq 0.1$, 
which leads to the result
\beq
M^{Pad\acute{e}}_{opt}(m''\to 0) \simeq 2.97\;\Lam(2)\;
\label{Mnum}
\eeq
for $n_f=2$. Similarly, we obtain $M^{Pad\acute{e}}_{opt}(m''\to 0) 
\simeq 2.85 \Lam(3)$
for $n_f=3$.
Note that these values of the dynamical quark masses, if 
they are to be consistent 
with the expected 
range~\cite{Miransky}
 of $M_{dyn}\simeq$ 300-400 GeV, call for relatively low
$\Lam \simeq $ 100-150 GeV, which 
is indeed supported by our results in the next section.
\section{Composite operators and $F_\pi$} 
We shall now generalize the ansatz 
(\ref{contour7})
for the pion decay constant $F_\pi$. 
The main idea is to
do perturbation theory around the same RG evolution solution 
with the non-trivial fixed point, as specified by the function $F$ in
(\ref{Fdef}),  
with perturbative correction terms obviously specific to $F_\pi$. 
A definition 
of $F_\pi$ suiting all our purposes is~\cite{GaLeut,derafael}
\beq
i\;\int d^4q e^{iq.x} \langle 0 \vert T\;A^i_\mu(x) A^k_\nu(0) \vert 0
\rangle =  
\delta^{ik} g_{\mu \nu} F^2_\pi +{\cal O}(p_\mu p_\nu)
\label{Fpidef}
\eeq
where the axial vector current $A^i_\mu \equiv 
 (\bar q \gamma_\mu
\gamma_5\lambda^i q)/2$ (the $\lambda^i$'s are Gell-Mann $SU(3)$ matrices
or Pauli matrices for $n_f =3$, $n_f=2$, respectively). Note that according
to (\ref{Fpidef}), $F_\pi$ is to be considered as an order
parameter of CSB~\cite{Stern}. \\

The perturbative expansion of (\ref{Fpidef}) for $m \neq 0$ is available
to the three-loop order, as it can be easily  
extracted
from the very similar contributions to the electroweak
$\rho$-parameter, calculated at two loops in \cite{abdel}
and three loops in \cite{Avdeev}.

The appropriate generalization of 
(\ref{contour7}) for $F_\pi$ 
now takes the form
\bea
& \ds{{F^2_\pi \over{\Lam^2}} = (2b_0)\; 
{2^{-2 C} a^2\over{2 i \pi}} \oint {dy\over y}\; y^2 {e^{y/m^{''}}}
\frac{1}{F^{\;2 A-1} [C + F]^{\;2 B}} }
\; \times \nn \\
& \ds{ {\delta_{\pi }
 \left(1 +{\alpha_{\pi}(a)\over{F}}+{\beta_{\pi}(a)
\over{F^2}}
\;+\cdots \right)} }
\label{Fpiansatz}
\eea
in terms of $F(y)$ defined by eq.~(\ref{Fdef}) and where 
$\delta_\pi$, $\alpha_\pi(1)$ and $\beta_\pi(1)$, 
whose complicated expressions
will be given elsewhere~\cite{jlk}, 
are fixed by matching the perturbative $\MS$
expansion in a way to be specified next.
Formula (\ref{Fpiansatz}) 
necessitates some comments: apart from the obvious changes in the powers of 
$F$, $y$, etc,  
dictated by dimensional analysis, 
note that the perturbative expansion of the (composite operator)
 $\langle A_\mu A_\nu \rangle$ in (\ref{Fpidef}) 
starts at one-loop, but zero $g^2$ order.
This leads to the extra $2b_0 F$ factor in (\ref{Fpiansatz}), 
corresponding to an expansion 
starting at ${\cal O}(1/g^2)$\footnote{ 
The ${\cal O}(1/g^2)$ 
first-order term cancels anyhow 
after the necessary subtraction discussed
below.}.
Another difference is that 
the perturbative expansion
of (\ref{Fpidef}) is ambiguous
due to remaining divergences
after mass and coupling 
constant renormalization. Accordingly it necessitates additional 
subtractions which, within our framework,  
are nothing but the usual 
renormalization procedure for a composite operator, which is (perturbatively)
well-defined~\cite{Collins}. 
The only consequence is that,  
after a consistent treatment
of the subtracted terms (i.e respecting RG invariance),
the unambiguous determination of the $1/F^n$ perturbative
terms in (\ref{Fpiansatz}) 
necessitates the knowledge of the $(n+1)$ order
of the ordinary perturbative expansion.
The nice thing, however, is that the subtracted terms only affect the
values of $\alpha_\pi$ and $\beta_\pi$, but not the
{\em form} of the ansatz (\ref{Fpiansatz}), as soon as the order of the 
variational-perturbative expansion is larger than 1~\cite{gn2}. 
The consistency of 
our formalism is checked by noting that the re-expansion of 
(\ref{Fpiansatz})
do reproduce correctly the LL and NLL dependence in $\bar m$ of the
perturbative expansion of the composite operator to all orders.

The 
analyticity range with respect to $\Delta\gamma_1$, discussed in section
2, remains valid for
(\ref{Fpiansatz}) 
as well, since the branch cuts are determined by the very same 
relation (\ref{Fdef}). We can thus proceed to a 
numerical optimization with respect to the RS dependence, along the
same line as the mass case in section 2. Using an appropriate Pad\'e
approximant form to define the $F \to 0$ ($m'' \to 0$) limit, 
we obtain the optimal 
values as 
\beq
F^{Pad\acute{e}}_{\pi ,opt}(m'' \to 0)
 \simeq 0.55\;\Lam(2)\;\;\;(0.59\;\Lam(3)\;)\;,
\eeq
for $n_f =$ 2 (3). With $F_\pi \simeq 92$ MeV,
this gives $\Lam \simeq $ 157 (168) MeV, for $n_f =$ 3 (2).
\section{$\langle \bar q q \rangle$ ansatz}
As is well known~\cite{Collins,Miransky}, 
$\langle \bar q q \rangle$ is not RG-invariant, while  
$m \langle \bar q q \rangle$ is; this is thus the relevant quantity 
to consider 
for applying our RG-invariant construction. 
A straightforward generalization of the 
derivation in section 3 leads to the ansatz 
\bea
{\bar m \langle \bar q q\rangle \over{\Lam^4}} =(2b_0)  
{2^{-4 C} a^4\over{2 i \pi}} \oint {dy\over y} {e^{y/m^{''}} y^4
\over{(F)^{4 A-1} [C + F]^{4 B
}}} {
\delta_{\langle \bar q q\rangle}
 \left(1 +{\alpha_{\langle \bar q q\rangle}(a)\over{F(y)}}
\;\right)}
\label{qqansatz}
\eea
where again the coefficients 
$\delta_{\langle\bar q q\rangle}$ and
$\alpha_{\langle\bar q q\rangle}(1)$ are obtained from matching the
ordinary perturbative expansion after a subtraction, 
and will be given explicitely elsewhere~\cite{jlk}. 
The perturbative expansion, known up to two-loop order 
\cite{Spiridonov,jlk}
implies that one only knows unambiguously the first
order correction ${\cal O}(1/F)$ in (\ref{qqansatz}), 
as previously discussed. 
Apart from that, (\ref{qqansatz})
has all the expected properties (RG invariance, resumming LL 
and NLL dependence etc),  
but a clear inconvenience 
is that $\langle\bar q q\rangle$ cannot be directly accessed, being  
screened by tiny explicit symmetry breaking effects due to $m \neq 0$. 
This is of course a well-known problem, not specific to our construction.
 
However, it is not clear how to consistently include explicit 
symmetry breaking effects
within our framework. 
As amply discussed, in (\ref{qqansatz}) 
$m^{''}$ is an arbitrary parameter, destined to reach its 
chiral limit $m^{''} \to 0$. 
Accordingly, $\bar m \to 0$
for $m'' \to 0$, 
so that one presumably expects only to recover a trivial result, 
$\bar m \langle\bar q q\rangle \to 0$ for $m'' \to 0$. This is 
actually the case:
although  
(\ref{qqansatz}) potentially gives a non-trivial result in the 
chiral limit, namely the simple
pole residue ($\equiv 2b_0(2C)^{-C} \;\delta_{\langle \bar q q\rangle}
\;\alpha_{\langle \bar q q\rangle}(a)$, 
upon neglecting unknown higher-order 
purely perturbative corrections), when we require 
extrema of 
this expression with respect to RS changes, 
using for the $m'' \to 0$ limit 
a Pad\'e approximant similar to the one for $F_\pi$,
we do {\em not} find 
non-zero extrema. 
Such a result is not conclusive regarding  
the actual value of $\langle\bar q q \rangle(\bar\mu)$, 
but it may be considered 
a consistency cross-check 
of our formalism. \\
 
On the other hand, we should mention that our basic expression
(\ref{qqansatz}) {\em does} possess non-trivial extrema for some 
$m''_{opt} \neq 0$. These 
we however refrain from interpreating in a more quantitative way
since, within our framework, we cannot 
give to $m''\times \Lam$ the meaning of a true, explicit quark mass 
(whose
input we in principle need in order to extract a $\langle \bar q q\rangle
$ value
from (\ref{qqansatz})).
At least, it strongly indicates that it should be possible 
to extract $\langle \bar q q\rangle$ in the chiral limit,  
by introducing in a consistent way  
a small explicit symmetry-breaking mass, 
$-m_{0,exp} \bar q_i q_i$, to the basic Lagrangian (\ref{xdef}). 
\section{Summary}
In this paper we have shown that 
the variational expansion in arbitrary $m''$, as developed
in the context of the GN model~\cite{gn2}, can be formally  
extended to the QCD case, apart from the complication due to the
presence of extra singularities, which can be however removed
by appropriate RS change. 
As a result we 
obtain in the chiral limit non-trivial relationships between  
$\Lam$ and the dynamical masses and order parameters, $F_\pi$,  
$\bar m \langle\bar q q\rangle$.
The resulting
expressions in a generalized RS have been numerically optimized, using 
a well-motivated Pad\'e approximant form, due to the complexity of the 
full optimization problem. The optimal values obtained for $M_q$ and $F_\pi$
are quite encouraging, while for $\langle\bar q q\rangle$ 
they are quantitatively
not conclusive,  
due to the inherent screening by an explicit
mass term of this quantity, in the limit $m \to 0$.
A possible extension to include consistently explicit breaking mass terms
in our formalism is explored in ref.~\cite{jlk}.
\vskip .5 cm
{\large \bf Acknowledgements} \\
We are grateful to Eduardo de Rafael for 
valuable remarks and discussions. 
J.-L. K. also thanks Georges Grunberg, Heinrich Leutwyler, Jan Stern
and Christof Wetterich for useful discussions. 
(C.A) is grateful to the theory group of Imperial College
for their hospitality.\\

\end{document}